# Cloud+: A safe and restrained data access control program for cloud

*Sarath Pathari*
*myappleid2849@gmail.com*

**ABSTRACT**

*Secure distributed storage, which is a rising cloud administration, is planned to guarantee the mystery of re-appropriated data yet also to give versatile data access to cloud customers whose data is out of physical control. Ciphertext-Policy Attribute-Based Encryption (CP-ABE) is seen as a champion among the most reassuring frameworks that may be used to verify the confirmation of the administration. Be that as it may, the use of CP-ABE may yield an unavoidable security burst which is known as the abuse of access accreditation (for example decoding right). In this paper, we look at the two essential occurrences of access accreditation misuse: one is on the semi-believed specialist side, and the other is supportive of the cloud customer. To ease the abuse, we propose revocable CP-ABE based distributed storage structure with express renouncing, planned information getting to and numerous examining capacities alluded as Cloud+.*

## 1. INTRODUCTION

The prevalence of distributed computing may in an indirect manner reason for the weakness of the security of re-appropriated information moreover, the insurance of cloud customers. A particular test here is on the most capable technique to guarantee that elite affirmed customers can get access to the information, which has been redistributed to the cloud, at wherever and at whatever point. One guileless course of action is to use an encryption framework on the information before transferring to the cloud. However, it limits information sharing plan. This is so in light of the way that an information owner needs to download the encoded information from the cloud and further re-scramble them for sharing (expect the information owner has no close-by copies of the information). A fine-grained get the opportunity to command over encoded information is appealing with respect to distributed computing.

Ciphertext-Policy Attribute-Based Encryption (CP-ABE) may be a successful response for guaranteeing the classification of information and to give fine-grained access control here. In a CP-ABE based distributed storage system, for example, affiliations and individuals would first be able to determine to get to strategy over properties of a potential cloud customer. Affirmed cloud customers by then are yielded get to accreditations (i.e., decoding keys) relating to their property sets which can be used to gain admittance to the redistributed information. CP-ABE offers a reliable technique to secure information set away in the cloud, yet in like manner engages fine-grained access authority over the information.

Any information that is secured in the cloud at whatever point spilled, could result in sort out of repercussions for the alliance and people. The current CP-ABE based [1] plan enables us to keep security crack from the outside attacker and besides an insider of the alliance who executes the "violations" of redistributing the unscrambling rights and the dissemination of understudy information in the plain game plan for illegal money related picks ups. Simultaneously, it can likewise guarantee that semi-believed expert won't (re-)circulate the made access certifications to others by proposing CryptCloud+, which gave a responsible specialist and revocable CP-ABE based distributed storage framework. Regardless, one attempting issue in dealing with customer repudiation in distributed storage is that a denied customer may regardless will, in any case, have the ability to unscramble an old ciphertext they were affirmed to access before being disavowed. To address this issue, the ciphertext set away in the circulated stockpiling ought to be invigorated, in a perfect world by the (untrusted) cloud server. In like manner, it required arranged data getting the chance to control which would give a noteworthy level of security.

This paper proposes Cloud+, an all-encompassing model of the CryptCloud+ by giving coordinated information access control. Furthermore, different evaluating and examination plot and evacuated one of the two revocable frameworks and diminished it to one express revocable framework. We have fused Shengmin's et. al [2] time encoding component into the revocable framework to give coordinated information access control. This paper broadens work in [1] as pursues.

(a) We address an inadequacy in the examining philosophy in [1]. Specifically, the inspecting approach will tumble for this situation just evaluator said to be trusted totally may disregard to be clear now and again. As a balance, we adjust the examining technique and add various evaluators summary to audit and consider results.
(b) We improve the value of the advancement of ATER-CP-ABE in [1]. This ATR-CP-ABE advancement empowers us to enough renounce the harmful customers explicitly.

## 2. RELATED WORK

In the paper [1] makers investigated the two essential examples of access capability misuse: one is on the semi-confined in the master side, and the other is agreeable to cloud customer. To direct the maltreatment, creators proposed the essential dependable expert and revocable CP-ABE based appropriated stockpiling system with white-box discernibility and checking on, implied as CryptCloud+.

In the paper [2], creators illustrated a capable revocable characteristic based encryption (RABE) plot with the property of ciphertext assignment. propose an ensured and successful fine-grained get the opportunity to control and data sharing arrangement for dynamic customer bundles by (1) portraying and maintaining access approaches in perspective on the qualities of the data; (2) permitting key age focus (KGC) to viably invigorate customer affirmations for dynamic customer get-togethers; and (3) allowing some exorbitant estimation tasks to be performed by untrusted CSPs without requiring any designation key.

In the paper [3], creator proposed the Secure Information Sharing in Clouds (SeDaSC) approach that gives data protection and trustworthiness, get the opportunity to control data sharing (sending) without using figure concentrated re-encryption, insider hazard security and forward and in turn around access control.

This paper [4] developed another cryptosystem for fine-grained sharing of mixed data. In this cryptosystem, ciphertexts are named with sets of qualities and private keys are connected with getting to structures that control which ciphertexts a customer is proficient to interpret.

This paper [5], proposed a multi-master ciphertext-approach ABE contrive with obligation, which licenses following the character of a getting into devilishness customer who discharged the translating key to other people, and as such lessens the trust in assumptions on the pros just as the customers.

This paper [6], proposes a thought called auditable σ-time redistributed CP-ABE, which is acknowledged to be fitting to dispersed processing. In this, an exorbitant mixing action brought about by translating is offloaded to the cloud and then, the rightness of the assignment can be assessed viably.

This paper [7], gave an expressive, gainful and revocable data get the chance to control plot for multi-master disseminated capacity systems, where there are different experts exist together and each authority can issue properties self-governing. Specifically, they proposed a revocable multi-master CP-ABE contrive, and connected it as the basic frameworks to diagram the information access to control plot.

## 3. METHODOLOGY
Our methodology of finding the noxious client and forestalling repudiated client from further getting to the ciphertext is demonstrated as follows.

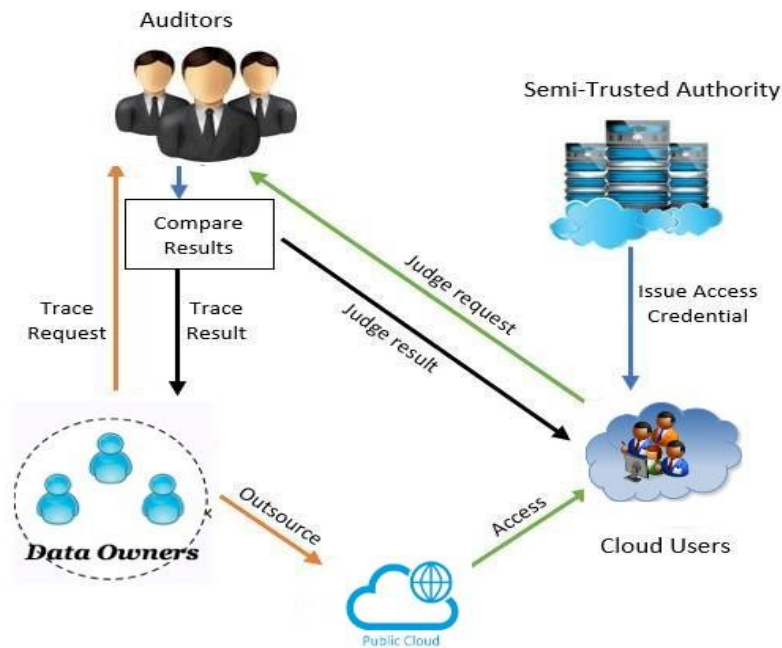

**Fig. 1: Architecture of Cloud+ System**

Right off the bat, we will depict the design of new Cloud++ information access control framework. Figure 1 above demonstrates the engineering of our framework. Framework contains five modules which are portrayed underneath.

(a) Data Owner: Data Owners (DOs) scrambles their data under the noteworthy access approaches before redistributing the (encoded) data to an open cloud (Public Cloud).
(b) Public Cloud Storage: An untrusted distributed storage specialist co-op where information redistributed by DO will be put away.
(c) Cloud Users: CU's are end-clients who solicitations for access to re-appropriated records.
(d) Auditors: They are the Third-Party reviewers who confirm, reviews, follows cloud clients and repudiates malignant clients of the cloud framework.
(e) Semi-Trusted Authority: They are in charge of setting up the framework, age of keys and redistribution of keys to approved and unapproved clients.

Semi-Trusted Authority is semi-trusted as in it may (re-)pass on access qualifications to the people who are unapproved anyway makes structure parameters (to be granted to Auditor) truly. A totally believed Auditor keeps a copy of the system parameters shared by Semi-Trusted Authority. Information Owners encode their data to deflect unapproved get to. Affirmed DUs may purposely discharge their passage certifications, for a model, pitching accreditations to an outcast. For all intents and purposes, get to accreditations are most likely going to pull in potential buyers (in dim grandstand), and the structure backstabbers (offering the affirmations) may never have been gotten.

In any event, one out of three will be an untrustworthy evaluator who unrealistically gives inaccurate review results to the information proprietors. So as to moderate this, we propose a different inspector plot whose free reviews are looked at and the right review results are sent to the approved information proprietors. These review results are thought about for greatest exactness in results. Furthermore, the review result with a legitimate review is sent back to the information proprietor.

We also revise the ATER-CP-ABE algorithm provided in paper [1] by incorporating time encoding algorithm provided by Shengmin et, al [2] in his paper. $sk_{-ten, id}$

Our ATR-CP-ABE has the following algorithms.
- **Setup**(*Lamda, U*) = (*Pp, msk*): Security parameter PS and universe of properties U is given as info which yields open parameters Pp and a Master mystery key msk. Likewise, it instates a vacant denial list RL.
- **TimeEncode**(*Date, T*) = $t_{en}$: Given the input *T*(bounded system life), *Date* (current date when the key is generated) we get bit string $t_{en}$ of the size $\log_2 T$ as output.
- **TimeDecode**($sk_{id,S,+ten}, cDate, cTime$) = $sk_{id,S,-ten}$. Given the info cDate(current date during unscrambling), cTime(current time during decoding), we get mystery key $sk_{id,S,-ten}$ without time component.
- **KeyGen**(*Pp, msk, id, S, $t_{en}$*) = $sk_{id, S,+ten}$: Key age stage between Semi confided in Authority and Users. We give open parameter Pp, ace mystery key msk and set of traits S with use character id alongside encoded time bits $t_{en}$. We get a Secret key $sk_{id, S,+ten}$ as output which is associated with the user id.
- **Encrypt**(*Pp, msg, AS, RL*) = *cipherText*: On the contribution of Pp, a plaintext message msg, an entrance structure AS over the universe of traits, and a renouncement list RL, it yields a cipherText.
- **Decrypt**(*Pp, $sk_{id,S,+ten}$, cipherText, cDate, cTime*) = *msg | error* : On input of public parameter *Pp*, a secret key $sk_{id, S,+ten}$, and a ciphertext *cipherText*, it first extracts time component of the secret key by performing **TimeDecode**($sk_{id, S,+ten}, cDate, cTime$) and outputs a secret key $sk_{id, S,-ten}$. At that point checks if the decoded time part is still not exactly current cDate and cTime.If assume the time segment is not exactly current date and time during unscrambling, at that point calculation continues to discover plaintext. It outputs the plaintext *msg* if the attribute set S of $sk_{-ten, id}$ satisfies the access structure of *cipherText* and *id* does not belong to *RL*. Otherwise, it outputs *error*.
- **KeyFormCheck**(*Pp, sk*) = 1 | 0 : 1=wellFormed 0= weaklyFormed. On input Pp and a secret key $sk_{id, S,+ten}$, it outputs 1 if $sk_{id, S,+ten}$ passes the key sanity check. Otherwise, it outputs 0.
- **Trace**(*Pp, msk, sk*) = $sk^*_{id}$ | noTraceReq: On information Pp, msk and a mystery key sk, it first checks whether sk is well-framed so as to further decide if sk should be followed. A mystery key sk characterized also framed if KeyFormCheck(Pp, sk)= 1. For an all around shaped sk, it removes the personality from sk. It at that point yields a character with which the sk partners, and places it in the renouncement list RL. Else, it yields a noTraceReq image showing that sk shouldn't be followed.
- **Audit**(*Pp, $sk_{id,S,+ten}$, $sk^*_{id}$*) = 1 | 0 1=guilty 0=innocent. If found guilty, the user is revoked and the decryption key is regenerated by performing KeyGen(*Pp, msk, id, S, $t_{en}$*) with new time encoded parameter.

Our Cloud+ works as follows:
- **System setup**: Semi-Trusted Authority arrangements the framework. It runs Setup(Lamda, U) = (Pp, msk)to produce framework open parameter Pp and ace mystery key msk. It shares this Pp and msk with Auditors before distributing Pp to other people.
- **User registration**:: A user gets successfully registered to the system and is issued an identity *id* an attribute set S is assigned to a user. Semi-Trusted Authority generates first generates an encoded time bits which specify the expiration of access to the data by calling TimeEncode(*Date, T*) = $t_{en}$. .He at that point creates a mystery get to accreditation for the client dependent on his character id and characteristic set S and joins time encoded bit strings to the scrambled access key by calling KeyGen(*Pp, msk, id, S, $t_{en}$*) = $sk_{id, S,+ten}$.
- **FileOutsource**: Data Owner first encrypts his file using AES data encryption technique and generates a session key *mskey* and defines access policy and encrypts the data by calling **Encrypt**(*Pp, $m_{skey}$, AS, RL*) = *cipherText*.

- **File Access**: Data user requests for the file access. Cloud returns requested file to the data user. Data user calls **Decrypt** (*Pp, $sk_{id}$, S, +ten, cipherText, cDate, cTime*) = *msg | error* to decrypt the encrypted data.
- **Trace**: When the auditor finds a secret access credential is being sold online or receives a trace request from a DataOwner, it runs **Trace**(*Pp, msk, sk*) = *$sk^*_{id}$ | noTraceReq* to find out who the leaker is.
- **Audit**: At the point when an information client with character id is followed as the leaker yet guarantees guiltlessness, it sends a review demand alongside his/her entrance certification to evaluators. After accepting the review demand, Auditors calls blameworthy or guiltless **Audit**(*Pp, $sk_{id}$, S, +ten, $sk^*_{id}$*) = 1 | 0 to determine whether the (accused) user is indeed guiltless, where access credential $sk^*_{id}$ is the leaked access credential.

## 4. EXPERIMENTAL RESULTS

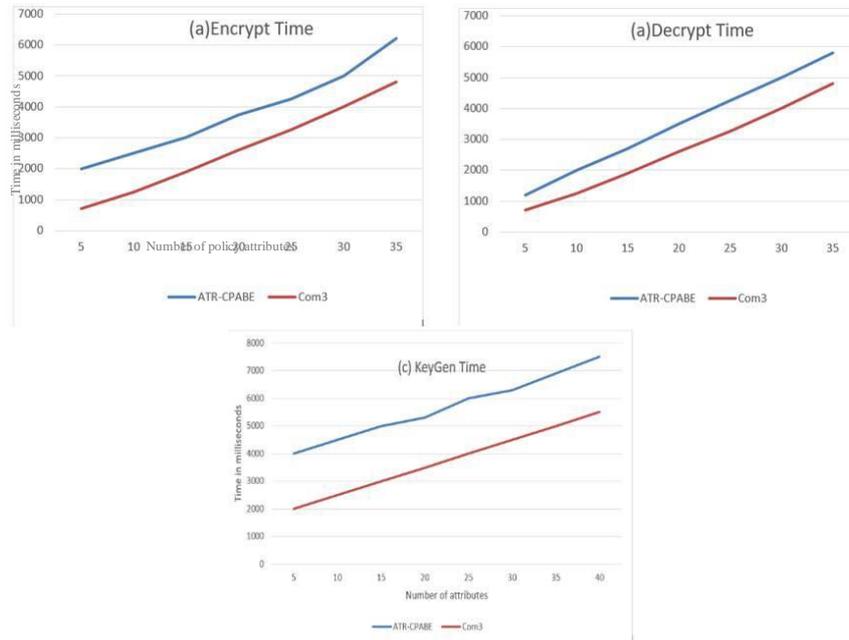

**Fig. 2: Experimental Results**

In CP-ABE systems, the multifaceted idea of ciphertext course of action impacts both the encryption time and the unscrambling time. To speak to this, we produce ciphertext approaches in the sort of (S1 and S2 ... likewise, Sl) to re-institute the most skeptical situation condition, where Si is an attribute. We try to survey the adequacy of ATR-CP-ABE taking a gander at the total time taken in the midst of each stage with the remarkable CP-ABE contrive which does not consider the passage accreditations misuse issue and the repudiation issue. As outlined in Figure 1, we take a gander at the time cost of executing a particular stage (tallying the Encrypt Time, the Decrypt Time and the KeyGen Time (of AT)). Since we consider both access capability misuse issue and the refusal issue, it isn't stunning to watch that our systems require extra time.

## 5. CONCLUSION

In this work, we have attempted to take care of the issue of certification spillage in CP-ABE based dispersed capacity structure by arranging a responsible specialist and revocable Cloud+. This is the CP-ABE based dispersed capacity structure that supports responsible expert, different reviewing and ground-breaking repudiation. In particular, Cloud++ empowers us to pursue and deny toxic cloud customers. Our methodology can be similarly used as a piece of the circumstance where the customers' confirmations are redistributed by the semi-put confidence in pro. This moreover gives a coordinated information access control where customer can get to the data inside a decided time for the given key thusly counteracting access to records by the denied client. Moreover, AU is believed to be totally trusted in CryptCloud+. Nevertheless, for all intents and purposes, it may not be the circumstance. So, we gave a way to deal with diminishing trust from AU by using different AU's.

## 6. REFERENCES

[1] "CryptCloud+: Secure and Expressive Data Access Control for Cloud Storage" Jianting Ning, Zhenfu Cao, Senior Member, IEEE, Xiaolei Dong, Kaitai Liang, Member, IEEE, Lifei Wei, and Kim-Kwang Raymond Choo, Senior Member, IEEE

[2] "Secure Fine-Grained Access Control and Data Sharing for Dynamic Groups in Cloud" Shengmin Xu, Guomin Yang, Senior Member, IEEE, Yi Mu, Senior Member, IEEE, and Robert H. Deng Fellow, IEEE

[3] Mazhar Ali, Revathi Dhamotharan, Eraj Khan, Samee U. Khan, Athanasios V. Vasilakos, Keqin Li, and Albert Y. Zomaya. "Sedasc: Secure data sharing in clouds." IEEE Vipul Goyal, Omkant Pandey, Amit Sahai, and Brent Waters. "Attribute-based encryption for fine-grained access control of encrypted data." In Proceedings of the 13th ACM conference on Computer and communications security, pages 89–98. ACM, 2006.

[4] Jin Li, Qiong Huang, Xiaofeng Chen, Sherman SM Chow, Duncan S Wong, and Dongqing Xie. "Multi-authority ciphertext-policy attribute-based encryption with accountability." In Proceedings of the 6th ACM Symposium on Information, Computer and Communications Security, ASIACCS 2011, pages 386–390. ACM, 2011.

[5] Jianting Ning, Zhenfu Cao, Xiaolei Dong, Kaitai Liang, Hui Ma, and LifeiWei. 'Auditable -time outsourced attribute-based

encryption for access control in cloud computing." IEEE Transactions on Information Forensics and Security,13(1):94–105, 2018.

[6] Kan Yang, Student Member, IEEE, and Xiaohua Jia, Fellow, IEEE. "Expressive, Efficient, and Revocable Data Access Control for Multi- Authority Cloud Storage."

[7] Amit Sahai and Brent Waters. "Fuzzy identity-based encryption." In Advances in Cryptology–EUROCRYPT 2005, pages 457–473. Springer, 2005.

[8] Brent Waters. "Ciphertext-policy attribute-based encryption: An expressive, efficient, and provably secure realization." In Public Key Cryptography–PKC 2011, pages 53–70. Springer, 2011.

[9] Kan Yang, Zhen Liu, Xiaohua Jia, and Xuemin Sherman Shen. "Time-domain attribute-based access control for cloud-based video content sharing: A cryptographic approach." IEEE Transactions on Multimedia, 18(5):940–950, 2016.

[10] Zhen Liu, Zhenfu Cao, and Duncan S Wong. "Blackbox traceable cp-abe: how to catch people leaking their keys by selling decryption devices on eBay." In Proceedings of the 2013 ACM SIGSAC conference on Computer & communications security pages 475–486. ACM, 2013Systems Journal, 11(2):395–404,2017.

[11] Zheng Yan, Xueyun Li, Mingjun Wang, and Athanasios V. Vasi- lakos. Flexible data access control based on trust and reputation in clo *loud Computing*, 5(3):485–498,2017